\documentclass{IEEEtran}
\usepackage{cite}
\usepackage{graphicx}
\usepackage{amsmath,amssymb,amsfonts}
\usepackage{subcaption}
\usepackage{textcomp}
\def\BibTeX{{\rm B\kern-.05em{\sc i\kern-.025em b}\kern-.08em
    T\kern-.1667em\lower.7ex\hbox{E}\kern-.125emX}}

\usepackage{algorithm}
\usepackage{algpseudocode}
\algrenewcommand\algorithmicrequire{\textbf{Input:}}
\algrenewcommand\algorithmicensure{\textbf{Output:}}
\newcommand{\Algheading}[1]{\Statex \textbf{#1}}
\newcommand{\blue}[1]{#1}

\newtheorem{definition}{Definition}
\newtheorem{theorem}{Theorem}
\newtheorem{lemma}{Lemma}

\newtheorem{remark}{Remark}
\newtheorem{example}{Example}
\title{Exact and Parametric Dynamical System Representation of Nonlinear Functions}
\author{Toshiyuki Ohtsuka, \IEEEmembership{Senior Member, IEEE}%
\thanks{This work was supported in part by JSPS KAKENHI Grant Number 26K23810.}%
\thanks{T. Ohtsuka is with the Department of Informatics, Graduate School of Informatics, Kyoto University, Kyoto, 606-8501 Japan. Email: ohtsuka@i.kyoto-u.ac.jp.}}

\begin{document}

\maketitle

\begin{abstract}
\blue{We introduce fixed-initial-state constant-input dynamical system (FISCIDS) representations for nonlinear functions. A function argument serves as a constant input to an input-affine system with a fixed initial state, and the function value is the terminal output. Every nonsingular differentially algebraic function on an open set star-shaped at the origin admits an exact quadratic FISCIDS representation. At least one analytic non-differentially-algebraic function admits such a representation. These results establish exact representation guarantees for structured finite-dimensional parametric model classes.}
\end{abstract}


\section{Introduction}

\blue{Function representations are fundamental tools in science and engineering. 
Common representation frameworks include polynomials~\cite{Dierckx94}, piecewise affine functions~\cite{Bemporad99}, orthogonal basis functions~\cite{Mallat99}, radial basis functions~\cite{Broomhead88,Poggio90}, Gaussian processes~\cite{Rasmussen06}, and neural networks~\cite{Bishop95,Haykin09}.} 
In recent years, deep neural networks (DNNs)~\cite{Lecun15,Goodfellow16} have been used in numerous applications in various fields. 

\blue{Residual networks \cite{He16}, a class of DNNs in which each layer updates a hidden state by adding a learned residual, can be interpreted as Euler discretizations of nonautonomous ordinary differential equations (ODEs).} 
This interpretation has motivated flow-based representations of nonlinear functions, namely representations of nonlinear functions using flows of dynamical systems~\cite{Haber17,Lu18,Chen18,Dupont19,Ruthotto20}. 
Flow-based representations are also used in flow-based generative models~\cite{Rezende15,Kingma18,Grathwohl19}.
\blue{Various expressivity and approximation results have been established for neural networks \cite{Cybenko89, Hornik89, Hornik91, Pinkus99} and, more recently, for ODE-based or flow-based architectures~\cite{Li23,Demarinis26}. These results motivate the use of dynamical systems for function representation.} 

\blue{In contrast to previous research focusing mainly on function approximation by dynamical systems, in this paper we consider a question about exact representation:  
\textit{Can dynamical systems of a simple structure exactly represent a broad class of nonlinear functions?}}   
This is not merely a question about flow-based representations or neural networks; it is a fundamental question about dynamical systems.  
\blue{A control-oriented motivation for structured representations arises from
the parameterization of nonlinear functions appearing in analysis, synthesis, and identification of dynamical systems. 
For example, in stability analysis and controller synthesis for nonlinear systems, Lyapunov functions and feedback laws are nonlinear functions of the state. 
Similarly, in nonlinear system identification, a one-step-ahead predictor or a state-transition map is a nonlinear function of measured states, inputs, and/or past outputs. 
A structured representation of such a function with a finite number of parameters is an essential tool in systems and control. 
The present paper does not develop algorithms for analysis, synthesis, or identification based on a structured parameterization; rather, it establishes exact representation guarantees for structured finite-dimensional model classes.}

This paper addresses the above question by introducing a dynamical system with a special structure: a \textit{fixed-initial-state constant-input dynamical system (FISCIDS) representation} of a nonlinear function. 
A FISCIDS representation is an input-affine dynamical system with a fixed initial state and a constant input \blue{(Definition \ref{def:FISCIDS})}. 
For a given nonlinear function, its argument is the constant input to its FISCIDS representation, \blue{whereas the initial state of the FISCIDS representation is fixed regardless of the argument.}
Then, the value of the nonlinear function is the output of the FISCIDS representation at a \textit{fixed terminal time}. 
We establish several fundamental properties and special classes of FISCIDS representations and their relationships to some classes of nonlinear functions, such as analytic functions and differentially algebraic (DA) functions \blue{(Definition \ref{def:DA})}. 

The contributions of this paper are summarized below. 
\begin{enumerate}
    \item We introduce a FISCIDS representation and its special forms\blue{:} rational, polynomial, and quadratic FISCIDS representations (Definition \ref{def:FISCIDS}). To the best of the author's knowledge, these representations for nonlinear functions have not been reported in the literature. 
    \item We establish several fundamental properties of FISCIDS representations (Theorems \ref{th:any}--\ref{th:equivalence}) \blue{and prove that any nonsingular DA function on an open set star-shaped at the origin} has a quadratic FISCIDS representation, which consists of at most quadratic polynomials (Theorem \ref{th:main}). 
    \item \blue{We formalize the construction of rational, polynomial, and quadratic FISCIDS representations from nonsingular DA certificates and give their dimension bounds (Algorithm \ref{alg:Construction}).} 
    \item We also show that there exists an analytic function that is not DA but has a quadratic FISCIDS representation (Theorem \ref{th:FISCIDS_TT}). 
    \blue{This implies that the set of quadratic-FISCIDS-representable functions contains an analytic function outside the set of DA functions.}  
\end{enumerate}
\blue{The class of DA functions contains many standard analytic functions used in science and engineering, including algebraic, exponential, trigonometric, and Bessel functions, and is closed under several common operations. 
Nonanalytic functions, such as discontinuous or nonsmooth functions, are outside the scope of this paper.}  

The remainder of this paper is organized as follows. 
Section~\ref{sec:DSR} introduces FISCIDS representations, 
Section~\ref{sec:DA} constructs rational, polynomial, and quadratic FISCIDS representations of nonsingular DA functions, and
Section~\ref{sec:TT} presents a Q-FISCIDS representation of an analytic non-DA function. 
Section~\ref{sec:Conclusion} concludes the paper.

\textit{Notation and Terminology:} 
For a vector $z \in \mathbb{R}^n$, $z_i$ denotes the $i$-th component of $z$. 
$\mathbb{Z}_{\ge 0}$ denotes the set of nonnegative integers. 
A set $X \subset \mathbb{R}^n$ is \textit{star-shaped at $\xi_0$} if $\xi_0 + s (\xi - \xi_0) \in X$ holds for all $\xi \in X$ and $s \in [0,1]$.

\section{Dynamical System Representation of Nonlinear Functions} \label{sec:DSR}

\subsection{Definition and Interpretation}\label{subsec:Def}
\blue{Hereafter, $\tau\in[0,1]$ denotes a fictitious time in a dynamical system for realizing a nonlinear function, and a dot over a variable denotes differentiation with respect to $\tau$. }

\begin{definition}[FISCIDS Representation] \label{def:FISCIDS}
Let $\phi: X \to \mathbb{R}^m$ be a function defined on an open set $X \subset \mathbb{R}^n$. 
Consider an $n$-input $m$-output $N$-dimensional input-affine dynamical system with a \textit{fixed initial state} $z_0 \in \mathbb{R}^N$ and a \textit{constant input} $\xi \in X$, which can be expressed as follows: 
\begin{align}
\Sigma \left\{
\begin{array}{l}
    \dot{z}(\tau) = F(z(\tau)) \xi, \quad z(0) = z_0, \\
    y(\tau) = h(z(\tau)), \\
\end{array}
\right. \label{eq:FISCIDS}
\end{align}
where $F : Z \to \mathbb{R}^{N \times n}$ and $h: Z \to \mathbb{R}^m$ are defined on an open set $Z \subset \mathbb{R}^N$ containing $z_0$. 
Let $z(\tau;\xi)$ and $y(\tau;\xi)$ be the state trajectory and output of $\Sigma$ for $\tau \ge 0$. 
If, for every $\xi \in X$, $z(\tau;\xi)$ is defined for all $\tau \in [0,1]$ and $y(1;\xi) = \phi(\xi)$ holds, $\Sigma$ is called a \textit{FISCIDS representation} of $\phi$. 
In particular, if $F(z)$ and $h(z)$ consist of rational functions, $\Sigma$ is called a \textit{rational FISCIDS (R-FISCIDS) representation} of $\phi$. 
If $F(z)$ and $h(z)$ consist of polynomials, $\Sigma$ is called a \textit{polynomial FISCIDS (P-FISCIDS) representation} of $\phi$. 
If $F(z)$ consists of at most quadratic polynomials and $h(z)$ consists of affine functions, $\Sigma$ is called a \textit{quadratic FISCIDS (Q-FISCIDS) representation} of $\phi$. 
\blue{A FISCIDS representation $\Sigma$ can also be specified by the tuple $(F(z),h(z),z_0)$.}
\end{definition}

\begin{remark}
    The constant input $\xi$ in (\ref{eq:FISCIDS}) can more generally be replaced by $\xi - \xi_0$ for a constant $\xi_0 \in \mathbb{R}^n$. 
    \blue{Moreover, the output map $h(z)$ does not need to be defined for the intermediate states $z(\tau;\xi)$ for $0 \le \tau <1$; it suffices that $h(z)$ is defined on an open set containing the terminal state set $z(1;X)$.} However, for simplicity, we set $\xi_0 = 0$ and assume that the output is defined whenever the state trajectory exists. 
\end{remark}

\blue{In this definition, ``fixed initial state'' means fixed with respect to the argument $\xi$ after a FISCIDS representation $(F(z),h(z),z_0)$ has been chosen. 
Once the state dimension and the rational or polynomial structure are fixed, the initial state $z_0$ and the coefficients in $F(z)$ and $h(z)$ form a finite-dimensional parameter vector that may depend on the given function $\phi(\xi)$. 
By contrast, $\xi$ is the argument at which the function is evaluated and varies from one evaluation to another. 
This distinction is important in applications: when parameterizing Lyapunov functions and feedback laws, $\xi$ may be the plant state; in system
identification, $\xi$ may be a regressor formed from measured inputs,
states, and outputs. Thus, the dynamics in a FISCIDS representation should be
regarded as a structured representation of a nonlinear function, not
necessarily as the physical dynamics of the plant.} 

\begin{figure*}
    \centering
    \subcaptionbox{Flow-based representation\label{fig:Flow}}{\includegraphics[height=4cm]{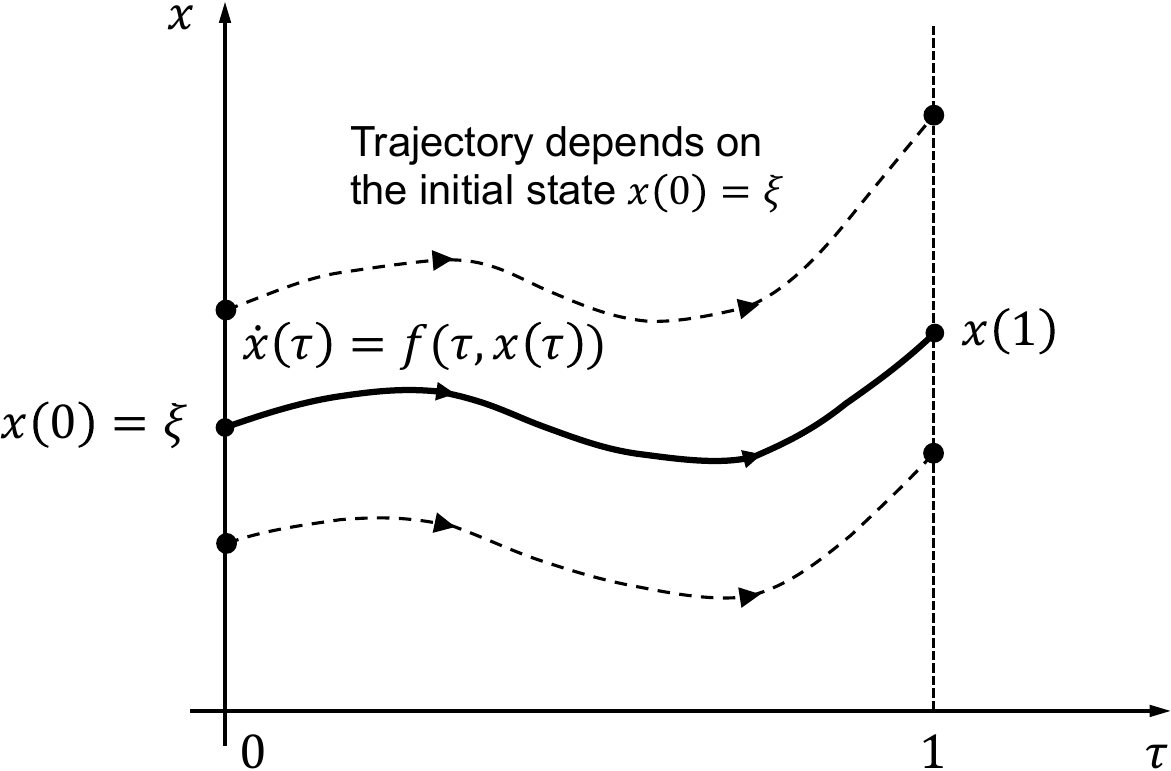}} %\\
    \hfil 
    \subcaptionbox{FISCIDS representation\label{fig:FISCIDS}}{\includegraphics[height=4cm]{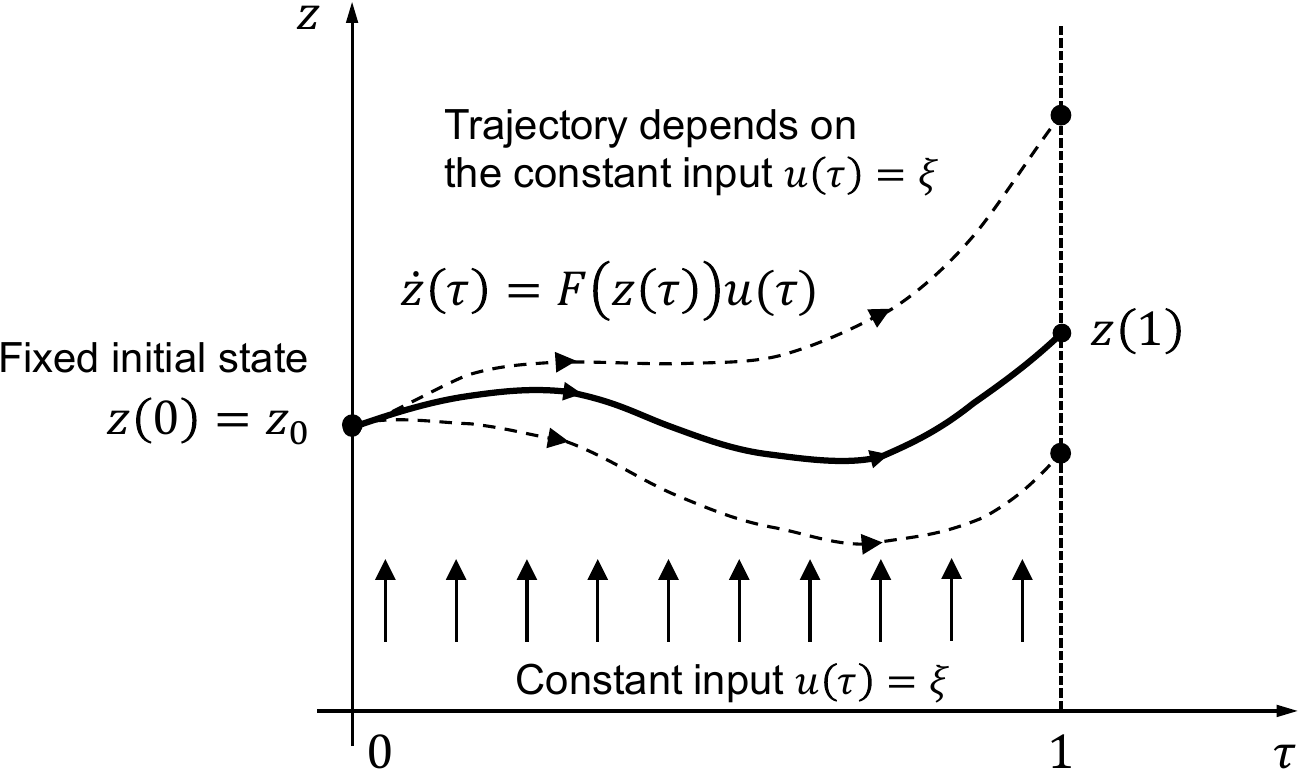}}
    \caption{Flow-based representation and FISCIDS representation of a nonlinear function. In a conventional flow-based representation (Fig.\ \ref{fig:Flow}), the argument of the nonlinear function determines the initial state of a possibly nonautonomous system. By contrast, in a FISCIDS representation (Fig.\ \ref{fig:FISCIDS}), the argument of the nonlinear function is \blue{the} constant input of an input-affine system with a fixed initial state. }
    \label{fig:TwoRepresentations}
\end{figure*}

A FISCIDS representation differs structurally from conventional flow-based representations, including neural ODEs \cite{Haber17,Lu18,Chen18,Dupont19,Ruthotto20}, in the way the function argument enters the dynamical system. 
As depicted in Fig.~\ref{fig:Flow}, in a conventional flow-based representation, the initial state of a possibly nonautonomous system depends on the argument of a nonlinear function. By contrast, in a FISCIDS representation, as shown in Fig.~\ref{fig:FISCIDS}, the initial state $z(0)$ of an input-affine dynamical system is \textit{fixed}, and the \textit{constant input} $u(\tau) = \xi$ $(0 \le \tau \le 1)$ is the argument of the nonlinear function. 
Every FISCIDS representation can nevertheless be written in the following autonomous flow-based form: 
\begin{align}
\left\{
\begin{array}{l}
\begin{bmatrix}
        \dot{z}(\tau) \\
        \dot{\zeta}(\tau)
    \end{bmatrix} 
    =
    \begin{bmatrix}
        F(z(\tau)) \zeta(\tau) \\
        0
    \end{bmatrix} , \quad
    \begin{bmatrix}
        z(0) \\
        \zeta(0)
    \end{bmatrix} 
    =
    \begin{bmatrix}
        z_0 \\
        \xi
    \end{bmatrix} , \\ 
    y(\tau) = h(z(\tau)). 
    \end{array}
    \right.  \label{eq:FBR}
\end{align}
\blue{In (\ref{eq:FBR}), the $z$-component of the initial state is fixed at $z_0$, whereas its $\zeta$-component equals the function argument $\xi$. 
The structural restriction that the argument enters only through the linear input $F(z)\xi$, while $z_0$ is independent of $\xi$, is preserved in the reduction of an R-FISCIDS representation to P- and Q-FISCIDS representations, as shown later in Theorem \ref{th:equivalence}.}  

Moreover, R-FISCIDS, P-FISCIDS, and Q-FISCIDS representations are different from classical notions of analog computability \cite{Shannon41} and system immersions into rational-in-the-state, polynomial-in-the-state, and quadratic-in-the-state representations\cite{Ohtsuka05} (or quadratization \cite{Bychkov24}) to simplify the structures of ODEs and dynamical systems. A Q-FISCIDS representation provides a function of a vector $\xi \in \mathbb{R}^n$, whereas analog computability deals with functions of a single variable. Moreover, a Q-FISCIDS representation depends only on a finite number of parameters with its initial state fixed. By contrast, a system immersion involves a nonlinear mapping of the initial state that is not necessarily represented by a finite number of parameters. 
R-FISCIDS representations are also different from universal differential equations \cite{Rubel81,Bournez20}, which have solutions that approximate any continuous univariate function to arbitrary accuracy. 
R-FISCIDS representations provide multivariate functions and do not necessarily generate outputs that approximate any continuous function. 

\subsection{Fundamental Properties}\label{subsec:Properties}
In this section, we establish three fundamental properties of FISCIDS representations in Theorems \ref{th:any}--\ref{th:equivalence}. 

\begin{theorem}[Trivial Exact FISCIDS Representation] \label{th:any}
Any function $\phi: X \to \mathbb{R}^m$ on an open set $X$ that is star-shaped at the origin has a FISCIDS representation, which is expressed as follows: 
\begin{align}
\Sigma \left\{
\begin{array}{l}
    \dot{z}(\tau) = \xi, \quad z(0) = 0, \\
    y(\tau) = \phi(z(\tau)). \\
\end{array}
\right. \label{eq:TrivialFISCIDS}
\end{align}
\end{theorem}

\begin{IEEEproof}
    It follows immediately that $z(\tau;\xi) = \xi \tau \in X$ \blue{for $\tau \in [0,1]$} and $y(1;\xi) = \phi(\xi)$ hold for all $\xi \in X$. 
\end{IEEEproof}

Equation (\ref{eq:TrivialFISCIDS}) is a \textit{trivial FISCIDS representation} of a function $\phi$.  
Although Theorem \ref{th:any} offers an exact representation of any function, (\ref{eq:TrivialFISCIDS}) is \textit{not a parametric representation} in general because it involves the function $\phi$ itself \blue{and does not define a fixed model structure that is independent of $\phi$. 
Accordingly, we focus on R-, P-, and Q-FISCIDS representations whose fixed-dimensional subclasses with prescribed algebraic structure form finite-dimensional parametric families.}

\begin{theorem}[Analyticity of R-FISCIDS-Representable Functions] \label{th:analyticity}
    If a function $\phi: X \to \mathbb{R}^m$ has an R-FISCIDS representation, it is analytic on $X$.
\end{theorem}

\begin{IEEEproof}
    Let the function $\phi$ have an R-FISCIDS representation \blue{$(F(z),h(z),z_0)$}, with $F(z)$ and $h(z)$ consisting of rational functions \blue{in} $z$. Then, the rational functions in $F(z)$ and $h(z)$ are analytic on an open set containing all trajectories $z(\tau;\xi)$ for $\xi \in X$ and $\tau \in [0,1]$. Therefore, owing to the analytic dependence of the solutions on the parameters (see \cite[(10.3)]{Lefschetz77}), $y(1;\xi) = \phi(\xi)$ is an analytic function of $\xi \in X$.  
\end{IEEEproof}

Although P-FISCIDS and Q-FISCIDS representations are special cases of R-FISCIDS representations, 
the following theorem shows that these representations are as general as R-FISCIDS representations. 

\begin{theorem}[Reduction from R-FISCIDS to P-FISCIDS and Q-FISCIDS Representations] \label{th:equivalence}
    If a function $\phi: X \to \mathbb{R}^m$ has an R-FISCIDS representation, it also has a P-FISCIDS representation and, in particular, a Q-FISCIDS representation. 
\end{theorem}

\begin{IEEEproof}
    We first construct a P-FISCIDS representation from an R-FISCIDS representation and then a Q-FISCIDS representation from a P-FISCIDS representation. 
    Let \blue{$(F(z),h(z),z_0)$} be an R-FISCIDS representation of a function $\phi: X \to \mathbb{R}^m$ with  $F(z)$ and $h(z)$ consisting of rational functions. Let $\{ d_1(z), \ldots, d_{M_R}(z) \}$ be the set of denominator polynomials in $F(z)$ and $h(z)$, and define $\tilde{\beta}(z) = [ 1/d_1(z), \ldots, 1/d_{M_R}(z) ]^{\top}$. 
    Since $F(z)$ and $h(z)$ are analytic on an open set $Z \blue{\subset} \mathbb{R}^N$ containing all trajectories $z(\tau;\xi)$ for $\xi \in X$ and $\tau \in [0,1]$, their denominator polynomials do not vanish and $\tilde{\beta}(z)$ is also analytic on $Z$. 

    The entries of $F(z)$ and $h(z)$ are represented as polynomials in $(z,\tilde{\beta}(z))$ and, therefore, there exist \blue{a matrix-valued polynomial map} $\tilde{F}(\tilde{z}_1,\tilde{z}_2)$ and \blue{a vector-valued polynomial map} $\tilde{h}(\tilde{z}_1,\tilde{z}_2)$ with $\tilde{z}_1 \in \mathbb{R}^{N}$ and $\tilde{z}_2 \in \mathbb{R}^{M_R}$ such that $F(z) = \tilde{F}(z,\tilde{\beta}(z))$ and $h(z) = \tilde{h}(z,\tilde{\beta}(z))$ hold for any $z \in Z$. 
    Moreover, for each element $\tilde{\beta}_i(z) = 1/d_i(z)$ of $\tilde{\beta}(z)$, we have 
    \begin{align*}
        \dfrac{\partial \tilde{\beta}_i(z)}{\partial z} = - \dfrac{1}{d_i(z)^2} \dfrac{\partial d_i(z)}{\partial z} = - \tilde{\beta}_i(z)^2 \dfrac{\partial d_i(z)}{\partial z},
    \end{align*}
    whose entries are again polynomials in $(z,\tilde{\beta}(z))$. Therefore, there exists a \blue{matrix-valued polynomial map}  $\tilde{G}(\tilde{z}_1,\tilde{z}_2)$ such that $\partial \tilde{\beta}(z) / \partial z = \tilde{G}(z,\tilde{\beta}(z))$. 

    \blue{Since} $\tilde{\beta}(z)$ is analytic on $Z$, $\tilde{z}_1(\tau) = z(\tau;\xi)$ and $\tilde{z}_2(\tau) = \tilde{\beta}(z(\tau;\xi))$ are defined for any state trajectory $z(\tau;\xi)$ of the R-FISCIDS representation $(F(z),h(z),z_0)$ for $\xi \in X$ and $\tau \in [0,1]$. 
    Moreover, \blue{by} using the \blue{matrix-valued polynomial map} $\tilde{F}(\tilde{z}_1,\tilde{z}_2)$, \blue{the vector-valued polynomial map} $\tilde{h}(\tilde{z}_1,\tilde{z}_2)$, and \blue{the matrix-valued polynomial map} $\tilde{G}(\tilde{z}_1,\tilde{z}_2)$ defined above, we have
    \begin{align*}
        \begin{bmatrix}
         \dot{\tilde{z}}_1 (\tau) \\
         \dot{\tilde{z}}_2 (\tau)     
        \end{bmatrix}
        &= 
        \begin{bmatrix}
         \dot{z}(\tau;\xi) \\
         \dfrac{\partial \tilde{\beta}(z(\tau;\xi))}{\partial z} \dot{z}(\tau;\xi)    
        \end{bmatrix} \notag \\
        &= 
        \begin{bmatrix}
         \tilde{F}(\tilde{z}_1(\tau),\tilde{z}_2(\tau)) \\
         \tilde{G}(\tilde{z}_1(\tau),\tilde{z}_2(\tau)) \tilde{F}(\tilde{z}_1(\tau),\tilde{z}_2(\tau))     
        \end{bmatrix} \xi , \\ %\displaybreak\\ 
        \begin{bmatrix}
         \tilde{z}_1 (0) \\
         \tilde{z}_2 (0)     
        \end{bmatrix}
        &= 
        \begin{bmatrix}
         z_0 \\
         \tilde{\beta}(z_0)    
        \end{bmatrix}, \\ 
        y(\tau;\xi) &= \tilde{h}(\tilde{z}_1(\tau),\tilde{z}_2(\tau)) ,
    \end{align*}
    which is a P-FISCIDS representation of $\phi(\xi)$. 
    Note that the initial state $(\tilde{z}_1(0),\tilde{z}_2(0))$ is fixed regardless of $\xi$ because $z_0 \in \mathbb{R}^N$ and $\tilde{\beta}(z_0) \in \mathbb{R}^{M_R}$ are fixed. 

    Next, consider a P-FISCIDS representation \blue{$(F_P(\tilde{z}),h_P(\tilde{z}),\tilde{z}_0)$} of $\phi(\xi)$ in the form 
    \begin{align*}
    \Sigma_P \left\{
    \begin{array}{l}
        \dot{\tilde{z}}(\tau) = F_P(\tilde{z}(\tau)) \xi, \quad \tilde{z}(0) = \tilde{z}_0 ,\\
        y(\tau) = h_P(\tilde{z}(\tau)) , 
        \end{array} \right.
    \end{align*}
    with $F_P: \tilde{Z} \to \mathbb{R}^{N_P \times n}$ and $h_P: \tilde{Z} \to \mathbb{R}^m$ whose entries are polynomials in $\tilde{z}$ on an open set $\tilde{Z} \subset  \mathbb{R}^{N_P}$, and $\tilde{z}_0 \in \tilde{Z}$. 
    Let $\mathcal{N}_{P} \subset \mathbb{Z}_{\ge 0}^{N_P}$ be the set of all multi-exponents of $\tilde{z}$ in $F_P(\tilde{z})$ and $h_P(\tilde{z})$, and let $\mathcal{M}_P$ be the set of corresponding monomials as $\mathcal{M}_P = \{ \tilde{z}^{\nu} : \nu \in \mathcal{N}_P \}$. 
    If either $F_P(\tilde{z})$ or $h_P(\tilde{z})$ contains a constant term, then $0 \in \mathcal{N}_P$ and $1 \in \mathcal{M}_P$. 
    Then, every entry of $F_P(\tilde{z})$ and $h_P(\tilde{z})$ belongs to $\operatorname{span} \mathcal{M}_P$, which is a linear subspace of polynomials over $\mathbb{R}$. 
    We define a set of multi-exponents
    \begin{align*}
        \bar{\mathcal{N}}_P = \left\{ \bar{\nu} \in \mathbb{Z}_{\ge 0}^{N_P} : \exists{\nu} \in \mathcal{N}_P \text{ \blue{such that} } \nu - \bar{\nu} \in \mathbb{Z}_{\ge 0}^{N_P} \right\}
    \end{align*}
    and the set of corresponding monomials as 
    \begin{align}
        \bar{\mathcal{M}}_P = \{ \tilde{z}^{\bar{\nu}} : \bar{\nu} \in \bar{\mathcal{N}}_P \}. \label{eq:AllMonomials}
    \end{align} 
    Then $\mathcal{N}_P \subset \bar{\mathcal{N}}_P$,  $\mathcal{M}_P \subset \bar{\mathcal{M}}_P$, and any partial derivative of any order of a monomial in $\mathcal{M}_P$ is a scalar multiple of a monomial in $\bar{\mathcal{M}}_P$. 
    Let $N_Q = |\bar{\mathcal{N}}_P|-1$ and let $\tilde{\gamma}(\tilde{z}) = [ \tilde{z}^{\bar{\nu}_1}, \ldots, \tilde{z}^{\bar{\nu}_{N_Q}} ]^\top$ be a vector of all monomials in $\bar{\mathcal{M}}_P \setminus \{1\}$. 
    The entries of $F_P(\tilde{z})$ and $h_P(\tilde{z})$ are represented as affine functions of $\tilde{\gamma}(\tilde{z})$, and\blue{,} therefore, there exist \blue{a matrix-valued affine map} $\bar{F}(\bar{z})$ and \blue{a vector-valued affine map} $\bar{h}(\bar{z})$ with $\bar{z} \in \mathbb{R}^{N_Q}$ such that $F_P(\tilde{z}) = \bar{F}(\tilde{\gamma}(\tilde{z}))$ and $h_P(\tilde{z}) = \bar{h}(\tilde{\gamma}(\tilde{z}))$ hold for any $\tilde{z} \in \tilde{Z}$. 
    Moreover, for each element $\tilde{\gamma}_i(\tilde{z}) = \tilde{z}^{\bar{\nu}_i}$ of $\tilde{\gamma}(\tilde{z})$, 
     $\partial \tilde{\gamma}_i(\tilde{z}) / \partial \tilde{z}_j$ is zero if $\bar{\nu}_{ij}=0$, and otherwise equals $\bar{\nu}_{ij} \tilde{z}^{\bar{\nu_i} - e_j} \in \operatorname{span} \bar{\mathcal{M}}_P$, 
    where $e_j \in \mathbb{Z}_{\ge 0}^{N_P}$ denotes the multi-exponent whose $j$-th component is one and whose other components are zero. 
    %Note that $\bar{\nu}_{ij} \beta^{\bar{\nu}_i - e_j}$ is zero when $\bar{\nu}_{ij} = 0$. 
    Since $\operatorname{span} \bar{\mathcal{M}}_P$ consists of affine functions of $\tilde{\gamma}(\tilde{z})$, there exists \blue{a matrix-valued affine map} $\bar{G}(\bar{z})$ such that $\partial \tilde{\gamma}(\tilde{z}) / \partial \tilde{z} = \bar{G}(\tilde{\gamma}(\tilde{z}))$. 

    Since the monomial vector $\tilde{\gamma}(\tilde{z})$ is analytic on $\tilde{Z}$, $\bar{z}(\tau) = \tilde{\gamma}(\tilde{z}(\tau;\xi))$ is defined for any state trajectory $\tilde{z}(\tau;\xi)$ of the P-FISCIDS representation for $\xi \in X$ and $\tau \in [0,1]$. 
    Moreover, \blue{by} using the \blue{matrix-valued affine map} $\bar{F}(\bar{z})$, \blue{the vector-valued affine map} $\bar{h}(\bar{z})$, and \blue{the matrix-valued affine map} $\bar{G}(\bar{z})$ defined above, we have
    \begin{align*}
         \dot{\bar{z}}(\tau)     
        &= 
         \dfrac{\partial \tilde{\gamma}(\tilde{z}(\tau;\xi))}{\partial \tilde{z}} \dot{\tilde{z}}(\tau;\xi)    
         = \bar{G}(\bar{z}(\tau)) \bar{F}(\bar{z}(\tau)) \xi , \\
         \bar{z}(0) &= \tilde{\gamma}(\tilde{z}_0) ,  \\ 
        y(\tau;\xi) &= \bar{h}(\bar{z}(\tau)) ,
    \end{align*}
    which is a Q-FISCIDS representation of $\phi(\xi)$. Note that the matrix $\bar{G}(\bar{z}) \bar{F}(\bar{z})$ is at most quadratic, and  
    the initial state $\bar{z}(0) = \tilde{\gamma}(\tilde{z}_0)$ is fixed regardless of $\xi$ because $\tilde{z}_0 \in \mathbb{R}^{N_P}$ is fixed. 
\end{IEEEproof}

\subsection{\blue{A Direct Illustration}}\label{subsec:Example1}
\blue{Before presenting the general construction algorithm, we give a simple
direct Q-FISCIDS realization to illustrate the basic mechanism of a
FISCIDS representation. The purpose of this example is to visualize how
a nonlinear function can be generated from a fixed initial state by
applying the argument of the function as a constant input. }

\begin{figure*}[tb]
\centering
\includegraphics[width=\linewidth]{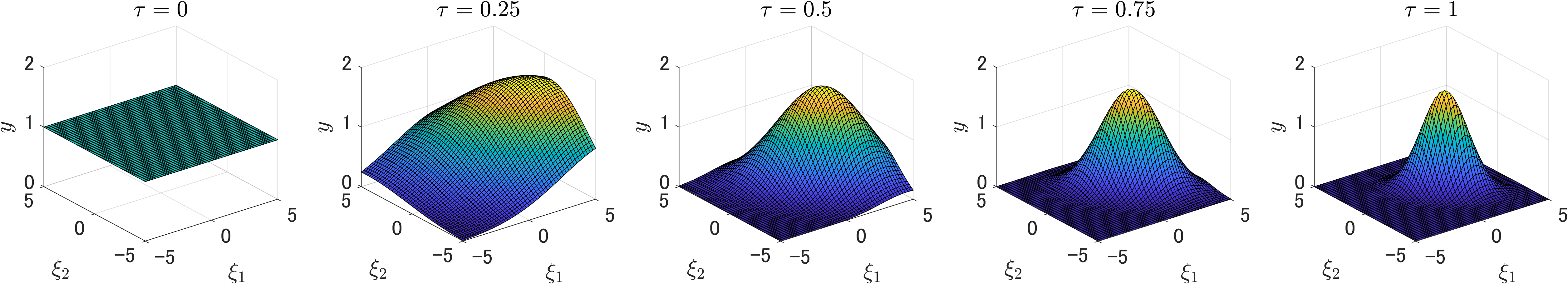}
\caption{Snapshots of the output $y(\tau;\xi)$ in Example~\ref{ex:Gaussian}. The Q-FISCIDS representation generates $y(1;\xi) = \exp \left({-\frac{1}{2}(\xi_1^2 + \xi_1 \xi_2 + \xi_2^2) + \xi_1 + \frac{1}{2}\xi_2} \right)$. The initial value $y(0;\xi) = 1$ is fixed for all $\xi$, \blue{whereas the terminal value $y(1;\xi)$ gives the target nonlinear function}.}
\label{fig:ExpQuad}
\end{figure*}

\begin{example}[A Direct Realization of a Gaussian-Type Function] \label{ex:Gaussian}
Consider a Gaussian-type function $\phi(\xi) = c \exp \left({\frac{1}{2}\xi^{\top}A \xi + b^{\top}\xi} \right)$, where $c \in \mathbb{R}$, $\xi, b \in \mathbb{R}^n$, and $A \in \mathbb{R}^{n \times n}$ is a symmetric matrix. This function admits a direct Q-FISCIDS realization as follows. 
We first define $\alpha_1(\xi) = \phi(\xi)$ and have 
    $\partial \alpha_1(\xi) / \partial \xi = \alpha_1(\xi) (\xi^{\top} A + b^{\top})$, which is bilinear in $\alpha_1(\xi)$ and $A \xi + b$.
    If we define $\alpha_2(\xi) = A \xi + b$, we have
    $\partial \alpha_2(\xi) / \partial \xi = A$. 
    Then, by defining $z(\tau) = [ z_1(\tau), z_2^{\top}(\tau) ]^{\top} = [ \alpha_1(\xi \tau), \alpha_2^{\top}(\xi \tau) ]^{\top}$, we obtain a Q-FISCIDS representation of $\phi(\xi)$ as
    \begin{align*}
    \left\{
    \begin{array}{ll}
    \blue{\begin{bmatrix}
        \dot{z}_1(\tau)\\
        \dot{z}_2(\tau)
    \end{bmatrix} = 
    \begin{bmatrix}
        z_1(\tau) z_2^{\top}(\tau) \\
        A
    \end{bmatrix} \xi, \quad
    \begin{bmatrix}
        z_1(0)\\
        z_2(0) 
    \end{bmatrix} = 
    \begin{bmatrix}
        \alpha_1(0)\\
        \alpha_2(0) 
    \end{bmatrix} = 
    \begin{bmatrix}
        c\\
        b 
    \end{bmatrix}}, \\
        y(\tau) = z_1(\tau) .
    \end{array} \right. 
    \end{align*}
    Figure \ref{fig:ExpQuad} shows snapshots of $y(\tau;\xi)$ for 
    \begin{align*}
        A = - \left[
        \begin{matrix}
            1 & 1/2 \\
            1/2 & 1 
        \end{matrix} \right], 
        b = \left[
        \begin{matrix}
            1 \\
            1/2  
        \end{matrix} \right], 
        c = 1. 
    \end{align*}
    The terminal value $y(1;\xi)$ coincides with $\phi(\xi)$, whereas the initial value $y(0;\xi) = c$ is fixed for all $\xi$. 
    \blue{This example highlights the separation between the fixed initial state
and the argument of the function. The initial state is determined by
the representation and is independent of $\xi$, whereas the trajectory
and the output vary with the constant input $\xi$. }
\end{example}

\section{FISCIDS Representations of DA Functions} \label{sec:DA}

\subsection{DA Functions and Exact Representation} \label{subsec:ExactRep}
\blue{In this section, we construct exact FISCIDS representations of DA functions \cite{Rubel85,Rubel89} that are nonsingular on open sets star-shaped at the origin. 
Lemma~\ref{lm:derivative-coordinate-closure} establishes a rational-closure property, which Theorem~\ref{th:main} uses to construct an R-FISCIDS representation. 
Theorem~\ref{th:equivalence} then yields P- and Q-FISCIDS representations. 
Algorithm~\ref{alg:Construction} and Example~\ref{ex:Log} summarize and illustrate the construction. }

\begin{definition}[DA Function] \label{def:DA}
    An analytic function $\phi: X \to \mathbb{R}$ defined on an open set $X \subset \mathbb{R}^n$ is called a \textit{DA function} if\blue{,} for each $i = 1, \ldots, n$, there exists a nonzero polynomial $p_i(\xi, Y_0, \ldots, Y_{\rho_i})$ with real coefficients and a nonnegative integer $\rho_i$ such that 
    \begin{align}
        p_i \!\left( \xi, \phi(\xi), \dfrac{\partial \phi(\xi)}{\partial \xi_i}, \ldots, \dfrac{\partial^{\rho_i} \phi(\xi)}{\partial \xi_i^{\rho_i}} \right) = 0   \label{eq:ADE}
    \end{align}
    holds for all $\xi \in X$. 
    Moreover, if 
    \begin{align}
        \dfrac{\partial p_i}{\partial Y_{\rho_i}} \!\left( \xi, \phi(\xi), \dfrac{\partial \phi(\xi)}{\partial \xi_i}, \ldots, \dfrac{\partial^{\rho_i} \phi(\xi)}{\partial \xi_i^{\rho_i}} \right) \neq 0  \label{eq:nonsigularity}
    \end{align}
    holds for all $\xi \in X$, $\phi$ is called \textit{nonsingular} on $X$. 
    A vector-valued function is called a nonsingular DA function if each component is DA and nonsingular. 
    If an analytic function is not a DA function, it is called a \textit{transcendentally transcendental (TT) function}. 
\end{definition}

For example, algebraic, exponential, trigonometric, and Bessel functions are DA functions of a single variable~\blue{\cite{Rubel89}}. 
Moreover, sums, products, quotients, compositions, and inverse functions (when defined) of DA functions are DA functions. 
\blue{Therefore, DA functions cover a large class of analytic functions commonly encountered in science and engineering.} 
However, there exist TT functions, such as Euler's gamma function and the Riemann zeta function~\cite{Rubel89}. 

Since DA functions satisfy algebraic differential equations (ADEs) in (\ref{eq:ADE}), they have the following strong property. 

\begin{lemma}[\blue{Rational Closure of Higher-Order Derivatives}]
\label{lm:derivative-coordinate-closure}
\blue{Let $\phi:X\to\mathbb{R}$ be a nonsingular DA function on an open set $X \subset\mathbb{R}^n$, satisfying (\ref{eq:ADE}) and (\ref{eq:nonsigularity}) for all $\xi \in X$. 
Define a set of partial derivatives
\begin{align}
\mathcal{A}_\phi = \left\{ \dfrac{\partial^{|\nu|}\phi(\xi)}{\partial \xi^\nu}: \nu \in \mathbb{Z}_{\geq0}^{n}, 0 \leq \nu_i \leq \rho_i, i=1,\ldots,n \right\}, \label{eq:set_derivatives}
\end{align}
where $|\nu| = \sum_{i=1}^{n} \nu_i$. 
Let $\alpha(\xi)$ be a vector of all elements of $\mathcal{A} = \{ \xi_1, \ldots, \xi_n \} \cup \mathcal{A}_\phi$, and let $N_R$ be the dimension of $\alpha(\xi)$. 
Then, for every component $\alpha_j(\xi)$ of $\alpha(\xi)$ and for every $k \in \{1,\ldots,n \}$, there exists a rational function $r_{jk}(z)$ of $z \in \mathbb{R}^{N_R}$ such that $r_{jk}(z)$ is analytic on an open set containing $\alpha(X)$ and satisfies
\begin{align*}
\dfrac{\partial \alpha_j(\xi)}{\partial \xi_k}
 = r_{jk}(\alpha(\xi)) \label{eq:finite-rational-closure}
\end{align*}
for all $\xi \in X$. 
Moreover, each $r_{jk}(z)$ admits a representation whose denominator is a power of the polynomial $\partial p_k/\partial Y_{\rho_k}$ in (\ref{eq:nonsigularity}) with $i=k$.}
\end{lemma}

\begin{IEEEproof}
\blue{We show that the partial derivative of every element $\alpha_j(\xi) \in \mathcal{A}$ with respect to $\xi_k$ is a rational function of $\alpha(\xi)$ with the specified form of denominator. 
Every element $\alpha_j(\xi) \in \mathcal{A}$ belongs to one of the following three cases. 
\begin{enumerate}
    \item If $\alpha_j(\xi) \in \{\xi_1, \ldots, \xi_n \}$, $\partial \alpha_j(\xi)/\partial \xi_k$ is either zero or one. 
    \item If $\alpha_j(\xi) = \partial^{|\nu|} \phi(\xi)/\partial \xi^{\nu} \in \mathcal{A}_{\phi}$ and $0 \le \nu_k < \rho_k$, then $\partial \alpha_j(\xi)/\partial \xi_k$ again belongs to $\mathcal{A}_{\phi}$ and is a component of $\alpha(\xi)$. 
    \item If $\alpha_j(\xi) = \partial^{|\nu|} \phi(\xi)/\partial \xi^{\nu} \in \mathcal{A}_{\phi}$ and $\nu_k = \rho_k$, then it is represented as
    \[ \alpha_j(\xi) = \dfrac{\partial^{|\nu'|}}{\partial \xi^{\nu'}} \left( \dfrac{\partial^{\rho_k} \phi(\xi)}{\partial \xi_k^{\rho_k}} \right), \]
where $\nu' = \nu - \rho_k e_k \in \mathbb{Z}_{\geq0}^{n}$ is identical to $\nu$ except that its $k$-th component is zero. 
    Therefore, $\alpha_j(\xi)$ is obtained by differentiating the highest-order derivative appearing in (\ref{eq:ADE}) for $i=k$. 
    In the following, we show that $\partial \alpha_j(\xi)/\partial \xi_k$ is a rational function of $\alpha(\xi)$. 
    By differentiating (\ref{eq:ADE}) for $i=k$ with respect to $\xi_k$, we have  
\begin{align*}
    \Gamma_{k}(\alpha(\xi)) + \sum_{\ell=0}^{\rho_k} \Delta_{k\ell}(\alpha(\xi)) \dfrac{\partial}{\partial \xi_k} \left( \dfrac{\partial^{\ell} \phi(\xi)}{\partial \xi_k^{\ell}} \right) = 0, 
\end{align*} 
for all $\xi \in X$, where
\begin{align*}
    \Gamma_{k}(\alpha(\xi)) &= \dfrac{\partial p_k}{\partial \xi_k}\!\left( \xi, \phi(\xi), \dfrac{\partial \phi(\xi)}{\partial \xi_k}, \ldots, \dfrac{\partial^{\rho_k} \phi(\xi)}{\partial \xi_k^{\rho_k}} \right), \\ %\notag \\
    \Delta_{k\ell}(\alpha(\xi)) &= \dfrac{\partial p_k}{\partial Y_{\ell}}\!\left( \xi, \phi(\xi), \dfrac{\partial \phi(\xi)}{\partial \xi_k}, \ldots, \dfrac{\partial^{\rho_k} \phi(\xi)}{\partial \xi_k^{\rho_k}} \right). %\label{eq:Delta}
\end{align*}
Since $\Delta_{k\rho_k}(\alpha(\xi)) \ne 0$ for all $\xi \in X$ by the nonsingularity assumption, the highest-order derivative can be represented as
\begin{align}
    &\dfrac{\partial}{\partial \xi_k} \left( \dfrac{\partial^{\rho_k} \phi(\xi)}{\partial \xi_k^{\rho_k}} \right) \notag \\ &= - \dfrac{\Gamma_{k}(\alpha(\xi)) + {\displaystyle \sum_{\ell=0}^{\rho_k-1}} \Delta_{k\ell}(\alpha(\xi)) \dfrac{\partial}{\partial \xi_k} \left( \dfrac{\partial^{\ell} \phi(\xi)}{\partial \xi_k^{\ell}} \right)}{\Delta_{k\rho_k}(\alpha(\xi))}. \label{eq:dalpha_rho_k}
\end{align} 
The right-hand side of (\ref{eq:dalpha_rho_k}) is a rational function of $\xi$ and $\partial^{\ell} \phi(\xi)/\partial \xi_k^{\ell}$ up to $\ell=\rho_k$. 
Since $\nu'_k=0$, the derivative 
\begin{align*}
\dfrac{\partial \alpha_j(\xi)}{\partial \xi_k} 
= \dfrac{\partial^{|\nu'|}}{\partial \xi^{\nu'}} \left( \dfrac{\partial}{\partial \xi_k}  \left( \dfrac{\partial^{\rho_k} \phi(\xi)}{\partial \xi_k^{\rho_k}} \right) \right),  
\end{align*}
introduces only mixed derivatives in $\mathcal{A}_{\phi}$, and it is a rational function of $\alpha(\xi)$ with its denominator consisting of a power of $\Delta_{k\rho_k}(\alpha(\xi))$. 
\end{enumerate}
The above three cases establish that the partial derivative of $\alpha_j(\xi)$ with respect to $\xi_k$ is a rational function of $\alpha(\xi)$ with its denominator consisting of a power of $\Delta_{k\rho_k}(\alpha(\xi))$. By the nonsingularity assumption and the continuity of a polynomial, $\Delta_{k\rho_k}(z)$ does not vanish on an open set containing $\alpha(X)$, as claimed. Note that $\Delta_{k\rho_k}$ is identical to the polynomial $\partial p_k/\partial Y_{\rho_k}$ in (\ref{eq:nonsigularity}) with $i=k$.}  
\end{IEEEproof}

\begin{theorem}[Exact FISCIDS Representations of Nonsingular DA Functions] \label{th:main}
    Any nonsingular DA function defined on an open set $X$ that is star-shaped at the origin has R-FISCIDS, P-FISCIDS, and Q-FISCIDS representations. 
\end{theorem}

\begin{IEEEproof}
    If every component of a vector-valued DA function $\phi$ has an R-FISCIDS representation, these R-FISCIDS representations are stacked to form an R-FISCIDS representation of $\phi$. 
    Therefore, it suffices to show the case where $\phi$ is scalar-valued. 
    We define $\alpha(\xi) = [\alpha_1(\xi), \ldots, \alpha_{N_R}(\xi)]^{\top}$ as a vector of all elements of $\mathcal{A} = \{\xi_1,\ldots,\xi_n\} \cup \mathcal{A}_{\phi}$, \blue{as defined in Lemma~\ref{lm:derivative-coordinate-closure}}. 
    Then, \blue{$\phi(\xi)$ is a component of $\alpha(\xi)$, and, for every $j \in \{1,\ldots,N_R\}$ and $k \in \{1,\ldots,n\}$, Lemma~\ref{lm:derivative-coordinate-closure} provides a rational function $r_{jk}(z)$ of $z \in \mathbb{R}^{N_R}$ satisfying $\partial \alpha_j(\xi)/\partial \xi_k = r_{jk}(\alpha(\xi))$. 
    If we define an $N_R \times n$ matrix-valued rational map $F_R(z) = [r_{jk}(z)]$ and a function $h_R(z) = z_{\ell}$ for $\ell$ such that $\phi(\xi) = \alpha_{\ell}(\xi)$,}  
    $\partial \alpha(\xi) / \partial \xi = F_R(\alpha(\xi))$ and $\phi(\xi) = h_R(\alpha(\xi))$ hold for all $\xi \in X$. 
    By nonsingularity, $F_R(z)$ and $h_R(z)$ are analytic on an open set containing $\alpha(X)$. 
    Then, for each $\xi \in X$, $z(\tau) = \alpha(\xi \tau)$ is defined for all $\tau \in [0,1]$ and satisfies 
    \begin{align*}
        \dot{z}(\tau) &= \dfrac{\partial \alpha(\xi \tau)}{\partial \xi} \dfrac{d (\xi \tau)}{d\tau} 
        = F_R(\alpha(\xi \tau)) \xi = F_R(z(\tau)) \xi, \\
        z(0) &= \alpha(0), 
    \end{align*}
    and $y(\tau) = h_R(z(\tau))$ 
    % \begin{align*}
    %     y(\tau) = h_R(z(\tau))
    % \end{align*}
    satisfies $y(1) = h_R(\alpha(\xi)) = \phi(\xi)$, 
    which is an R-FISCIDS representation of $\phi(\xi)$. 
    By Theorem \ref{th:equivalence}, $\phi(\xi)$ also has P-FISCIDS and Q-FISCIDS representations. 
\end{IEEEproof}

\blue{Thus, under the assumptions of Theorem~\ref{th:main}, every nonsingular DA function admits an exact Q-FISCIDS representation.} 

\subsection{\blue{Construction Algorithm and Dimension Bounds}} \label{subsec:Algorithm}
\blue{Under the assumptions of Theorem~\ref{th:main}, the construction of R-, P-, and Q-FISCIDS representations is explicit once a nonsingular DA certificate is given. 
Here, a DA certificate consists of the ADEs in Definition~\ref{def:DA}, their orders $\rho_i$, and the values of finite-order derivatives at the origin that specify a branch of the target function. 
Algorithm~\ref{alg:Construction} summarizes the construction steps of R-, P-, and Q-FISCIDS representations. 
Although it is stated for scalar-valued functions for notational simplicity, the vector-valued case is obtained by stacking the componentwise constructions. 
It is a certificate-to-realization procedure, and determining such a certificate or minimal orders $\rho_i$ is outside its scope.}

\begin{algorithm}[tbh]
\caption{\blue{Construction of R-FISCIDS, P-FISCIDS, and Q-FISCIDS Representations from a Nonsingular DA Certificate}}\label{alg:Construction}
\blue{\small \begin{algorithmic}[1]
\Require
A nonsingular DA certificate $(p_i, \rho_i)_{i=1}^n$ for $\phi$, including the values at the origin of the derivatives in $\mathcal{A}_{\phi}$ 
\Ensure
R-FISCIDS representation $(F_R(z), h_R(z), z_0)$, P-FISCIDS representation $(F_P(\tilde{z}), h_P(\tilde{z}), \tilde{z}_0)$, and Q-FISCIDS representation $(F_Q(\bar{z}), h_Q(\bar{z}), \bar{z}_0)$ of $\phi$
\Algheading{R-FISCIDS Representation:}
\State $\mathcal{A} \gets \{\xi_1, \ldots, \xi_n\} \cup \mathcal{A}_{\phi}$
\State $\alpha(\xi) \gets \text{an ordered vector of elements in }\mathcal{A}$
\State $F_R(z) \gets [r_{jk}(z)] \text{ in Lemma \ref{lm:derivative-coordinate-closure}}$
\State $h_R(z) \gets z_{\ell} \text{ for $\ell$ such that } \alpha_{\ell}(\xi) = \phi(\xi)$
\State $z_0 \gets \alpha(0)$
\Algheading{P-FISCIDS Representation:}
\State $\tilde{\beta}(z) \gets \text{a vector of the distinct nonconstant reciprocals}$ 
\Statex \qquad \qquad $1/\Delta_{k\rho_k}(z)\ (k=1,\ldots,n) \text{ in Lemma \ref{lm:derivative-coordinate-closure}}$ 
\State $\beta(z) \gets \left[z^{\top}, \tilde{\beta}^{\top}(z)\right]^{\top}$
\State $F_P(\tilde{z}) \gets \text{the matrix-valued polynomial map }$ 
\Statex \qquad \qquad $\left[ \tilde{F}^{\top}(\tilde{z}), (\tilde{G}(\tilde{z})\tilde{F}(\tilde{z}))^{\top} \right]^{\top} \text{ in Theorem \ref{th:equivalence}}$
\State $h_P(\tilde{z}) \gets \tilde{z}_{\ell}$
\State $\tilde{z}_0 \gets \beta(z_0)$ 
\Algheading{Q-FISCIDS Representation:}
\State $\bar{\mathcal{M}}_P \gets \text{the monomial set defined in (\ref{eq:AllMonomials})}$  
\State $\tilde{\gamma}(\tilde{z}) \gets \text{an ordered vector of elements in } \bar{\mathcal{M}}_P \setminus \{1\}$
\State $F_Q(\bar{z}) \gets \text{the matrix-valued quadratic map } \bar{G}(\bar{z}) \bar{F}(\bar{z})$
\Statex \qquad \qquad $\text{ in Theorem \ref{th:equivalence}}$ 
\State $h_Q(\bar{z}) \gets \bar{z}_m \text{ for $m$ such that } \tilde{\gamma}_m(\tilde{z}) = \tilde{z}_{\ell}$
\State $\bar{z}_0 \gets \tilde{\gamma}(\tilde{z}_0)$
\end{algorithmic}}
\end{algorithm}

\blue{These construction steps are similar to transformation steps used in existing methods for different problems} \cite{Shannon41,Ohtsuka05,Bychkov24}. 
\blue{A crucial difference between FISCIDS representations and existing methods is that the initial state is fixed.}  
Another difference is that the existence of trajectories is guaranteed for a fixed interval $[0,1]$. 

\blue{The respective dimensions $N_R$, $N_P$, and $N_Q$ of the R-, \mbox{P-,} and Q-FISCIDS representations produced by Algorithm~\ref{alg:Construction} satisfy
\begin{align*}
N_R &\le n+\prod_{i=1}^{n}(\rho_i+1), \quad
N_P  = N_R + M_R \le N_R + n, \\
N_Q &=|\overline{\mathcal{N}}_P|-1 \le \binom{N_P + d_P}{d_P} - 1,
\end{align*}
where $M_R$ is the number of reciprocal variables introduced by
$\tilde{\beta}(z)$, $\overline{\mathcal{N}}_P$ is the componentwise
downward closure of the multi-exponents of the monomials occurring in
$F_P(\tilde{z})$ and $h_P(\tilde{z})$, and $d_P$ denotes the maximum total degree of the entries of $F_P(\tilde{z})$ and $h_P(\tilde{z})$. 
As shown in Lemma \ref{lm:derivative-coordinate-closure}, the denominators in $F_R(z)$ are powers of the polynomials $\Delta_{k\rho_k}(z)$ $(k=1,\ldots,n)$. Therefore, $M_R$ is bounded by $n$. 
The maximum number of monomials in an $N_P$-variable polynomial of total degree $d_P$ is given by $\binom{N_P + d_P}{d_P}$. 
These formulas provide constructive upper bounds, and the resulting dimensions are not necessarily minimal. 
The bound on $N_R$ depends only on the input dimension $n$ and the orders $\rho_i$, whereas $N_P$ and $N_Q$ also depend on the denominator and monomial structures, and $N_Q$ may grow combinatorially. 
Thus, Algorithm~\ref{alg:Construction} provides an explicit construction, but it is not claimed to be computationally efficient.}

\blue{For fixed input and output dimensions and a fixed dimension $N$, Q-FISCIDS representations define a
finite-dimensional parametric family of nonlinear functions: 
\begin{align*}
\mathcal{Q}_N = 
\left\{
\xi\mapsto h_\theta(z_\theta(1;\xi)):
\dot z_\theta(\tau)=F_\theta(z_\theta(\tau))\xi,\ \allowbreak
z_\theta(0)=z_{0,\theta}
\right\},
\end{align*}
where $\theta$ consists of the coefficients of the matrix-valued quadratic map $F_\theta(z)$, the vector-valued affine map $h_\theta(z)$, and the initial state $z_{0,\theta}$. 
Theorem~\ref{th:main} guarantees that, for every nonsingular DA function on an open set star-shaped at the origin, there exists a dimension $N$ and a parameter vector $\theta$ such that the function belongs to $\mathcal Q_N$. 
Thus, whenever a Lyapunov function, feedback law, or predictor map satisfies the assumptions of Theorem \ref{th:main}, it belongs to $\mathcal{Q}_N$ for some $N$ without representation error. Corresponding analysis, synthesis, and identification algorithms are outside the scope of this paper.}

\subsection{\blue{An Algorithmic Example}}\label{subsec:Example2}

\begin{example}[\blue{Algorithmic Construction for a Logarithmic DA Function}] \label{ex:Log} 
\blue{This example illustrates the construction of R-, P-, and Q-FISCIDS representations from a nonsingular DA certificate. Consider $\phi(\xi) = \log P(\xi)$, where $\xi \in \mathbb{R}$ and $P(\xi) = a_0 + a_1 \xi + a_2 \xi^2$ $(a_0 > 0)$. 
On any open interval containing the origin on which $P(\xi)>0$, this function is a nonsingular DA function. The construction below follows Algorithm~\ref{alg:Construction}, with redundant coordinates suppressed for simplicity, and the dimensions may be below the general bounds. } 

\blue{\paragraph{R-FISCIDS Representation}
    We first specify the DA certificate. The function $\phi(\xi)$ satisfies the following ADE: 
    \[ p_1(\xi,\phi(\xi),\phi'(\xi)) = P(\xi) \phi'(\xi) - P'(\xi) = 0, \]
    for $p_1(\xi,Y_0,Y_1) = P(\xi) Y_1 - P'(\xi)$ and $\rho_1 =1$, where a prime denotes differentiation with respect to $\xi$. 
    Then, the sets $\mathcal{A}_{\phi}$ and $\mathcal{A}$ in Algorithm~\ref{alg:Construction} are given as $\mathcal{A}_{\phi} = \{\phi(\xi),\phi'(\xi)\}$ and $\mathcal{A} = \{\xi,\phi(\xi),\phi'(\xi)\}$, respectively. 
    The derivative of each element in $\mathcal{A}$ is a rational function of the elements of $\mathcal{A}$. 
    In particular, differentiating the ADE gives $\phi''(\xi) = - (P'(\xi) \phi'(\xi) - P''(\xi))/P(\xi)$, and $P(\xi)$ corresponds to $\Delta_{1\rho_1}$ in Lemma \ref{lm:derivative-coordinate-closure}.  
    However, in this example, $\phi'(\xi) = P'(\xi)/P(\xi)$ is already a rational function of $\xi$, and $\alpha(\xi) = [\xi, \phi(\xi)]^{\top}$ suffices to construct an R-FISCIDS representation. 
    In fact, we have $\alpha'(\xi) = [1, P'(\xi)/P(\xi) ]^{\top}$, and an R-FISCIDS representation $(F_R(z),h_R(z),z_0)$ is given by
    \begin{align*}
    \left\{
    \begin{array}{ll}
        \dot{z}(\tau) = F_R(z(\tau)) \xi ,  \quad z(0) = \alpha(0) = [0, \log a_0]^{\top}, \\
        y(\tau) = h_R(z(\tau)),
    \end{array} \right. \\
    F_R(z) = [1, P'(z_1)/P(z_1)]^{\top}, \quad h_R(z) = z_2.
    \end{align*}
One can confirm that $z(\tau) = \alpha(\xi \tau)$ is a trajectory of the R-FISCIDS representation, and $y(\tau) = z_2(\tau) = \phi(\xi \tau)$ holds. }

\blue{\paragraph{P-FISCIDS Representation}
Since the only denominator in the R-FISCIDS representation is $P(z_1)$, $\tilde{\beta}(z)$ in Algorithm~\ref{alg:Construction} is given by $1/P(z_1)$. 
Then, $\beta(z) = [z^{\top}, 1/P(z_1)]^{\top}$, and we have
\begin{align*}
\dfrac{\partial \beta(z)}{\partial z} = 
\begin{bmatrix}
    1 & 0 \\
    0 & 1 \\
  - \frac{P'(z_1)}{P(z_1)^2} & 0
\end{bmatrix}
= 
\begin{bmatrix}
    1 & 0 \\
    0 & 1 \\
  - P'(\beta_1(z))\beta_3(z)^2 & 0
\end{bmatrix}.
\end{align*}
Then, Algorithm~\ref{alg:Construction} yields a P-FISCIDS representation $(F_P(\tilde{z}),h_P(\tilde{z}),\tilde{z}_0)$ as follows: 
    \begin{align*}
    \left\{
    \begin{array}{ll}
        \dot{\tilde{z}}(\tau) = F_P(\tilde{z}(\tau)) \xi , %\\
        \quad \tilde{z}(0) = \beta(z_0) = [0,\log a_0,1/a_0]^{\top}, \\
        y(\tau) = h_P(\tilde{z}(\tau)) , 
    \end{array} \right. \\
    F_P(\tilde{z}) = 
    \begin{bmatrix}
        1 \\ (a_1 + 2a_2 \tilde{z}_1) \tilde{z}_3 \\ - (a_1 + 2a_2 \tilde{z}_1) \tilde{z}_3^2
    \end{bmatrix}, h_P(\tilde{z}) = \tilde{z}_2 ,
    \end{align*}
    whose entries are polynomials of degree at most three in $\tilde{z}$. 
    One can confirm that $\tilde{z}(\tau) = \beta(z(\tau)) = \beta(\alpha(\xi \tau))$ is a trajectory of the P-FISCIDS representation, and $y(\tau) = \tilde{z}_2(\tau) = \phi(\xi \tau)$ holds. }
    
\blue{\paragraph{Q-FISCIDS Representation}
The polynomial vector field above contains a cubic term. 
To apply the final stage of Algorithm~\ref{alg:Construction} and reduce the polynomial degree to two, we define the set of monomials in the P-FISCIDS representation as $\mathcal{M}_P = \{1, \tilde{z}_2, \tilde{z}_3, \tilde{z}_1 \tilde{z}_3, \tilde{z}_3^2, \tilde{z}_1 \tilde{z}_3^2 \}$. The partial derivatives of the monomials in $\mathcal{M}_P$ are scalar multiples of monomials in $\bar{\mathcal{M}}_P = \{1, \tilde{z}_1, \tilde{z}_2, \tilde{z}_3, \tilde{z}_1 \tilde{z}_3, \tilde{z}_3^2, \tilde{z}_1 \tilde{z}_3^2 \}$. 
    Then, $\tilde{\gamma}(\tilde{z})$ in Algorithm~\ref{alg:Construction} is given as $\tilde{\gamma}(\tilde{z}) = [ \tilde{z}_1, \tilde{z}_2, \tilde{z}_3, \tilde{z}_1 \tilde{z}_3, \tilde{z}_3^2, \tilde{z}_1 \tilde{z}_3^2 ]^{\top}$, and $F_P(\tilde{z})$, $h_P(\tilde{z})$, and $\partial \tilde{\gamma}(\tilde{z}) / \partial \tilde{z}$ are affine in $\tilde{\gamma}(\tilde{z})$. 
    Therefore, $(\partial \tilde{\gamma}(\tilde{z}) / \partial \tilde{z}) F_P(\tilde{z})$ is quadratic in $\tilde{\gamma}(\tilde{z})$, and Algorithm~\ref{alg:Construction} yields a Q-FISCIDS representation $(F_Q(\bar{z}),h_Q(\bar{z}),\bar{z}_0)$ as 
    \begin{align*}
    \left\{
    \begin{array}{ll}
        \dot{\bar{z}}(\tau) = F_Q(\bar{z}(\tau)) \xi , \\
        \bar{z}(0) = \tilde{\gamma}(\tilde{z}_0) = \left[0, \log a_0, 1/a_0, 0, 1/a_0^2, 0  \right]^{\top},\\
        y(\tau) = h_Q(\bar{z}(\tau)) ,
    \end{array} \right. \\
    F_Q(\bar{z}) = 
    \begin{bmatrix}
        1 \\ 
        a_1 \bar{z}_3 + 2a_2 \bar{z}_4 \\
        - a_1 \bar{z}_5 - 2a_2 \bar{z}_6 \\ 
        \bar{z}_3 - a_1 \bar{z}_1 \bar{z}_5 - 2a_2 \bar{z}_1 \bar{z}_6 \\
        -2 a_1 \bar{z}_3 \bar{z}_5 - 4a_2 \bar{z}_3 \bar{z}_6 \\
        \bar{z}_5 -2 a_1 \bar{z}_4 \bar{z}_5 - 4a_2 \bar{z}_4 \bar{z}_6
    \end{bmatrix}, \quad 
    h_Q(\bar{z}) = \bar{z}_2 . 
    \end{align*}
One can confirm that $\bar{z}(\tau) = \tilde{\gamma}(\tilde{z}(\tau)) = \tilde{\gamma} \circ \beta \circ \alpha(\xi \tau)$ is a trajectory of the Q-FISCIDS representation, and $y(\tau) = \bar{z}_2(\tau) = \phi(\xi \tau)$ holds. It should also be noted that the initial states $z(0)$, $\tilde{z}(0)$, and $\bar{z}(0)$ of the R-, P-, and Q-FISCIDS representations are fixed values and independent of the argument $\xi$.}
\end{example}

\section{Q-FISCIDS Representation of a TT Function} \label{sec:TT}

\blue{Under the assumptions of Theorem~\ref{th:main}, every nonsingular DA function has a Q-FISCIDS representation. Here, we show that the class of Q-FISCIDS-representable functions also contains at least one TT function. Hence, this class strictly contains the class of nonsingular DA functions covered by Theorem~\ref{th:main}. This does not imply that every TT function admits a Q-FISCIDS representation.} 
\blue{Since} no general characterization of TT functions is available, we present an example of a TT function having a Q-FISCIDS representation. 
This TT function is constructed by introducing parameters into Katriel's example \cite{Katriel03} of a differential equation with non-DA dependence on initial conditions. 
We modify Katriel's argument concerning dependence on initial values because a FISCIDS representation has a constant input and a fixed initial state. 

\begin{theorem}[Existence of a Q-FISCIDS-Representable TT Function]\label{th:FISCIDS_TT} 
There exists a TT function that has a Q-FISCIDS representation. 
\end{theorem}

\begin{IEEEproof}
    This is a direct consequence of the following Lemma \ref{lm:TT_example}. 
\end{IEEEproof}

\begin{lemma}[A Q-FISCIDS-Representable TT Function]\label{lm:TT_example}
    Consider the following system with constant inputs $\xi_1, \xi_2 \in \mathbb{R}$:
\begin{align} \left\{
    \begin{array}{ll}
        \dot{z}_1(\tau) = (z_2(\tau) - z_1(\tau)) \xi_1, & z_1(0) = 0, \\
        \dot{z}_2(\tau) = z_2(\tau) (z_2(\tau) - z_1(\tau)) \xi_2, & z_2(0) = 1, \\
        y(\tau) = z_1(\tau). 
    \end{array} \right. \label{eq:TT}
    \end{align}
    The output $y(1; \xi_1, \xi_2)$ is a TT function of $(\xi_1,\xi_2)$ on $\hat{X} = \{ (\xi_1,\xi_2) : 0 < \xi_1 < e \xi_2, 0 < \xi_2 < 1 \}$. 
    Thus, (\ref{eq:TT}) is a Q-FISCIDS representation of a TT function. 
\end{lemma}

\begin{IEEEproof}
    From (\ref{eq:TT}), we have $\dot{z}_2(\tau) / z_2(\tau) = \xi_2 \dot{z}_1(\tau) / \xi_1$, which leads to $z_2(\tau) = e^{\xi_2 z_1(\tau)/\xi_1}$ and $\dot{z}_1(\tau) = (e^{\xi_2 z_1(\tau)/\xi_1} - z_1(\tau)) \xi_1$. \blue{The latter equality can be rearranged to 
    \begin{align}
        \dfrac{\xi_2 \dot{z}_1(\tau) / \xi_1}{(\xi_2 / \xi_1) e^{\xi_2 z_1(\tau)/\xi_1} - \xi_2 z_1(\tau)/ \xi_1} = \xi_1 .  \label{eq:dF}
    \end{align}
    By integrating (\ref{eq:dF}) from $0$ to $\tau \ge 0$ with respect to the integration variable $\sigma$, 
    we have
    \begin{align*}
        \int_0^{\tau} \dfrac{\xi_2 \dot{z}_1(\sigma) / \xi_1}{(\xi_2 / \xi_1) e^{\xi_2 z_1(\sigma)/\xi_1} - \xi_2 z_1(\sigma)/ \xi_1} d\sigma = \xi_1 \tau. 
    \end{align*}
    By using the change of variable $v = \xi_2 z_1(\sigma)/\xi_1$ and $z_1(0)=0$, we obtain}
    \begin{align}
        F(\xi_2 z_1(\tau)/ \xi_1, \xi_2 / \xi_1) = \xi_1 \tau, \label{eq:Feq}
    \end{align}
    with \blue{the} function $F$ defined as
    \[ F(s,x) = \int_0^s \dfrac{dv}{x e^v - v} . \]
    The domain of definition $D_F$ of $F(s,x)$ contains $\{ (s,x) : s \in \mathbb{R}, x > 1/e \}$, and $F(s,x)$ is a TT function of $x$ for generic $s$ \cite{Katriel03}. 
    We denote the inverse function of $w = F(s,x)$ with respect to $s$ as $s = G(w,x)$. 
    By definition, $s = G(F(s,x),x)$ holds for any $(s,x) \in D_F$. 
    The domain of definition of $G(w,x)$ contains $\hat{D}_G = \{ (w,x) : 0 < w < 1/x, x > 1/e \}$, and $G(w,x)$ is a TT function of $x$ for generic $w$ \cite{Katriel03}. 
    \blue{Using (\ref{eq:Feq}) in the first argument of $G(w,\xi_2 / \xi_1)$, we obtain} $\xi_2 z_1(\tau)/ \xi_1 = G(\xi_1 \tau, \xi_2 / \xi_1)$. 
    Therefore, the output $y(\tau) = z_1(\tau)$ at $\tau=1$ is given by $y(1; \xi_1, \xi_2) = \xi_1 G(\xi_1, \xi_2 / \xi_1) / \xi_2$, \blue{which is shown to be a TT function on $\hat{X}$ by contradiction, as follows. 
    Suppose, to the contrary, that $y(1;\xi_1,\xi_2)$ is DA on $\hat{X}$. Choose a generic $w \in (0,e)$ for which $G(w,x)$ is TT in $x$. 
    Since DA functions are closed under substitution, $x \mapsto y(1;w,wx)$ is DA on $(1/e,1/w)$. Hence, $G(w,x)=x y(1;w,wx)$ is also DA there, which is a contradiction. }
\end{IEEEproof}

\section{Conclusion} \label{sec:Conclusion}
In this paper, \blue{we introduced FISCIDS representations and their special forms: R-FISCIDS, P-FISCIDS, and Q-FISCIDS representations, as exact dynamical-system representations of nonlinear functions. 
For fixed dimensions and prescribed algebraic structures, these forms define finite-dimensional parametric families. 
We established several fundamental properties and presented a construction algorithm to obtain these representations.}  
In particular, \blue{any nonsingular DA function on an open set star-shaped at the origin has a Q-FISCIDS representation, and at least one TT function also has such a representation.}  
\blue{Accordingly, these results suggest the use of Q-FISCIDS representations as structured finite-dimensional parameterizations of nonlinear functions such as Lyapunov functions, nonlinear feedback laws, and one-step-ahead predictors. Developing algorithms for analysis, synthesis, and identification of dynamical systems based on these parameterizations is left for future research. }

\section*{Acknowledgment}
The author would like to thank Kenta Hoshino for discussions that motivated this study.

\bibliographystyle{IEEEtran}
\bibliography{IEEEabrv,ODErep}

\end{document}